\documentclass[draft]{agujournal2019}
\usepackage{url} 
\usepackage{lineno}
\usepackage[inline]{trackchanges} 
\usepackage{soul}
\justifying
\draftfalse

\journalname{JGR: Space Physics}

\begin{document}

\title{A database of MMS bow shock crossings compiled using machine learning}

\authors{A. Lalti\affil{1,2}, Yu. V. Khotyaintsev\affil{1}, A. P. Dimmock\affil{1}, A. Johlander\affil{1}, D. B. Graham\affil{1}, V. Olshevsky.\affil{3}}

\affiliation{1}{Swedish Institute of Space Physics, Uppsala, Sweden}
\affiliation{2}{Space and Plasma Physics, Department of Physics and Astronomy, Uppsala University, Uppsala, Sweden}

\affiliation{3}{Main Astronomical Observatory, Kyiv, Ukraine}

\correspondingauthor{Ahmad Lalti}{ahmadl@irfu.se}

\begin{keypoints}
\item We use a machine learning technique to identify shock crossing events using the MMS spacecraft.
\item We compile the to date largest database of bow shock crossings with 2797 events including key parameters for each event.
\item Using the database we show that quasi-parallel shocks are more efficient at accelerating ions than their quasi-perpendicular counterpart.
\end{keypoints}

\begin{abstract}
Identifying collisionless shock crossings in data sent from spacecraft has so far been done manually. It is a tedious job that shock physicists have to go through if they want to conduct case studies or perform statistical studies. We use a machine learning approach to automatically identify shock crossings from the Magnetospheric Multiscale (MMS) spacecraft. We compile a database of those crossings including various spacecraft related and shock related parameters for each event. Furthermore, we show that the shocks in the database have properties that are spread out both in real space and parameter space. We also present a possible science application of the database by looking for correlations between ion acceleration efficiency at shocks and different shock parameters such as $\theta_{Bn}$ and $M_A$. Furthermore, we investigate statistically the ion acceleration efficiency. We find no clear correlation between the acceleration efficiency and $M_A$ and we find that quasi-parallel shocks are more efficient at accelerating ions. 
\end{abstract}


\section{Introduction}

    Collisionless shock research has occupied scientists for the last 70 years, they are one of the most dynamically rich phenomena in space. This stems from their non-linearity and their strong dependence on parameters such as the Mach number $M$, the angle between the shock normal vector and the upstream magnetic field $\theta_{Bn}$ and the plasma beta $\beta$. Shocks can be found across diverse plasma environments throughout the universe, from supernova remnants to interstellar and interplanetary media to planets. Despite the existence of extensive literature on the matter, the physics that dictates the evolution and dynamics of collisionless shocks is not fully understood. Many open questions remain, such as the different wave-particle processes in the shock ramp leading to the irreversible dissipation of solar wind bulk energy into heat, or the mechanisms that make collisionless shocks one of the most efficient particle accelerators in the universe \cite{treumann2009,bykov2011}. More theoretical, numerical, and observational work is required to be able to fully understand the physics of collisionless shock.
    
    In situ observations have played a major role in advancing our knowledge about collisionless shocks. The first spacecraft to cross Earth's bow shock and hence provide the first conclusive evidence for its existence was the Imp I spacecraft in 1964 (\citeA{ness1964initial}). After that, many spacecraft have been launched, equipped with instrumentation on board to study the space plasma environment of the solar system. Measurements by those spacecraft has allowed the investigation of collsionless shocks at various locations in the solar system such as interplanetary (IP) shocks in the solar wind \cite{kilpua2015properties} and bow shocks at non-terrestrial  planets\cite{sulaiman2016characterization,zhang2008initial}.
    
    With each new spacecraft more advanced instrumentation have been implemented, and new discoveries are made. To identify times when a spacecraft crosses the bow shock, vast amounts of data need to be surveyed by visually checking for characteristics of shock crossings. This is further complicated by  the search for shocks with restricted parameters that suit a science question of interest. Although reliable, this method is highly time-consuming. A multitude of space missions are presently, or have historically, encountered shock waves throughout the heliosphere. Current data archives amount to hundreds of thousands of hours of data to go through searching for shock crossings. This laborious manual task can be averted by the development of an automated approach to finding shocks in the data. From this approach, a database can be compiled containing the time and location of each shock crossing, along with the main parameters characterizing each shock. Such a database, would be a sizable asset to the space physics community that can help advance the knowledge of the physics of collisionless shocks.

    One of the most recent, and arguably the most advanced space mission with the purpose of studying space plasma physics is the Magnetospheric Multiscale (MMS) mission \cite{MMS_overview}. It was launched in 2015, and is a constellation of 4 spacecraft in a tetrahedral formation orbiting Earth equipped with high resolution fields and particle instruments with the primary goal of exploring electron-scale physics related to magnetic reconnection around Earth. Such high-resolution instrumentation has also proved to be extremely valuable for studying Earth's bow shock. The MMS spacecraft send around 16 gigabits of data per day \cite{MMS_overview} containing both fields and plasma measurement. Over the 6 years of operation, several terabytes of data is available for analysis. Many studies have used this data to investigate some of the still standing questions in collisionless shock physics, from particle acceleration \cite{amano2020observational,hanson2020shock}, to identifying different electrostatic and electromagnetic wave modes \cite{goodrich2018mms,hull2020mms,vasko2020nature}, to shock non-stationarity \cite{johlander2016rippled,yang2020mms,madanian2021dynamics}, with many more studies still expected to come.
    The high-cadence measurements by MMS and the years of data available present an ideal opportunity to compile an extensive bow shock database as discussed above.

    The most comprehensive shock crossings database for IP shocks, to our knowledge, is the University of Helsinki's Heliospheric Shock Database (www.ipshocks.fi/) where they provide, along with the time of crossing of the shock, other parameters that are necessary for understanding shock dynamics, such as the shock geometry, and the Alfv\'{e}nic Mach number ($M_A$ ). As for the Earth's bow shock, many databases of shock crossings have been compiled from different spacecraft. An example of a terrestrial bow shock database is that using observations by Imp 2 and 3 or the ISEE spacecraft, which is available at NASA Space Science Data Coordinated (NSSDC) archive. Furthermore, \citeA{kruparova2019statistical} compiled a database of 529 shock crossings using the Cluster spacecraft with a focus on studying the statistical dependence of the shock velocity on different parameters. 
    
    In this report we present a database of shock crossings observed by MMS, that was compiled using a machine learning technique \cite{olshevsky2021automated} for the detection of bow shock crossings. The database contains 2797 events along with key shock parameters. 
    We present the method used for identification of the shock crossings in Section \ref{section:2}. In Section \ref{section:3} we present the different parameters contained in the database and then discuss the uncertainties, caveats and drawbacks that one should keep in mind while using the database. In Section \ref{section:4} we present some examples of different shock crossings in the database, along with various statistical results highlighting the distribution of the shocks both in parameter space and in real space around Earth. To demonstrate the possible applications of the database, we perform a statistical study of the ion acceleration efficiency. Finally, in Section \ref{section:5} we summarize our results and state the conclusions.

\section{Automated Identification of Bow Shock Crossings}
\label{section:2}

In recent years machine learning algorithms have been extensively applied for data mining in various fields including space physics. Recently \citeA{olshevsky2021automated} have implemented a machine-learning algorithm to classify the different regions in space that MMS crosses throughout its orbit. MMS's orbit brings the spacecraft to four main plasma regions: undisturbed solar wind, solar wind with shock-reflected ions called the ion foreshock, magnetosheath, and magnetosphere. Each of those regions has characteristic signatures in the ion velocity distribution function (VDF).  \citeA{olshevsky2021automated} took advantage of the 3D VDFs measured by the Fast Plasma Investigation (FPI) ion instrument measuring at $4.5$ seconds resolution \cite{pollock2016fast} on MMS and trained a 3D Convolutional Neural Network (CNN) to identify the region in space where MMS is. For each ion VDF measurement, the CNN assigns a probability for the MMS to be in one of the four different regions, with the highest probability corresponding to the actual region in space for more than $98 \%$ of the time.

\begin{figure*}[ht]
    \centering
    \includegraphics[scale=0.3]{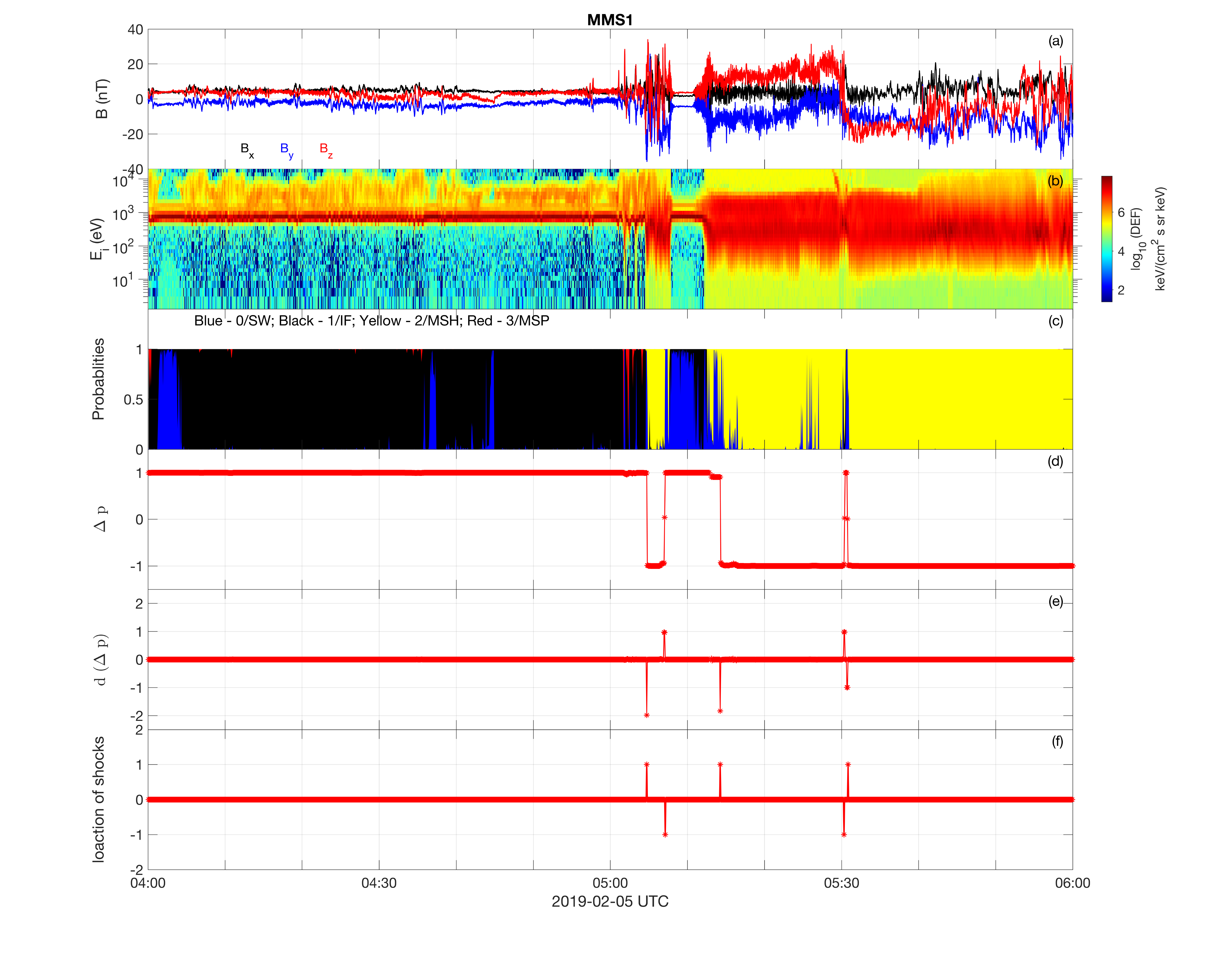}
    \caption{Shock crossing identification from MMS observations. (a) Magnetic field in GSE, (b) Omni-directional ion differential energy flux, (c) probability output from the CNN color coded with blue representing solar wind, black representing ion foreshock, yellow representing magentosheath and red representing magnetosphere, (d) probability difference at each time step $\Delta p (t)$, (e) difference of the probability difference $d\left(\Delta p (t)\right)$, and (f) detected shock crossings with 1 and -1 representing inbound and outbound crossings respectively.}
    \label{fig:fig1}
\end{figure*}

Using the identification of plasma regions by \citeA{olshevsky2021automated}, it's possible to identify when MMS traverses from the solar wind or foreshock into the magnetosheath, or vice versa and hence determine the shock crossing times. This is illustrated in Figure~\ref{fig:fig1}. Panel (a) shows the magnetic field, panel (b) shows the omnidirectional ion differential energy flux, panel (c) shows the probability output from the CNN color coded with blue representing solar wind, black representing ion foreshock, yellow representing magnetosheath and red representing magnetosphere. To determine the occurrence of shock crossings we calculate the probability difference at each measurement point
\begin{equation}
    \Delta p = (p_{SW} + p_{IF}) - p_{MSH},
\end{equation}
with $p_{SW}$,  $p_{IF}$, and  $p_{MSH}$ are the probabilities of the measurement being in the solar wind (SW), the ion foreshock (IF) and the magnetosheath (MSH) respectively. This quantity is shown in panel(d). If MMS is in the SW or IF $\Delta p \sim 1$ as in the region between 04:00 and 04:40 UT, while if MMS is in the MSH $\Delta p \sim -1$ as in the region between 05:40 and 06:00 UT. 
In mixed regions where the CNN was not able to specify with high confidence what region MMS is in $\Delta p$ will be noisy and fluctuate significantly and hence prevent accurate determination of shock transitions. To avoid this problem we put a threshold on $\Delta p$ where we remove all data with $|\Delta p| <0.9$. On top of that, we apply a moving median to $\Delta p$ to smooth it out, with varying window size, to detect shock transitions with different speeds. The window size varies from 2 to 50 measurement points or 9 to 225 seconds intervals. $\Delta p$ shown in panel (d) has a moving median applied to it with a 12 point window size. Then, to detect the time of transition, we calculate $\mathrm{d} (\Delta p)_i = \Delta p_{i+1}-\Delta p_i$, where i corresponds to the index of the current probability difference and $i+1$ corresponds to the next time step. This quantity is shown in panel (e), and at shock transitions this quantity should exhibit local maxima or minima depending if the shock is traversed from downstream to upstream (outbound), or vice versa (inbound), respectively.  The times for the extrema in $d (\Delta p (t))$ are identified as the shock transition time. Panel (f) shows the detected shock crossings with a value of 1 and -1 depending if it was an outbound or an inbound crossing.

From panels (a-b) we can see three different shock crossings in the interval between 05:00 and 05:20 UT. Comparing this with panels (c-f) we see that this method works well in identifying shock crossing events.

From time to time, the CNN mislabels one region for another, which could result in a misidentification of a shock crossing. An example of that is seen in Figure \ref{fig:fig1} around 05:30 UT where panels (a--b) show a magnetosheath current sheet. The CNN mistakenly labeled this region as a crossing from magnetosheath to ion foreshock and hence detecting an inbound and an outbound crossing of a shock. We visually check each shock and filter out such misidentifications from the final database. Finally, because of the variable window that we use with the moving median, the location of the shock crossing could be shifted to the upstream or the downstream, so we manually correct it to where there is a foot/ramp signature in the data.

\section{Compiling the Shock Database}
\label{section:3}

 In the final database, a total of 2797 shock crossings have been identified using the approach described above, spanning a period from October 2015 to December 2020. For the database to be of more use to scientists, we include various parameters that are essential for understanding collisionless shocks physics; all of which are described below and shown in Table \ref{tab:signif_gap_clos}. One can categorize the parameters into two groups, ones that relate to the spacecraft and data acquisition mode, and the others related to shock crossing itself.

We start by describing the parameters relating to the shock crossing itself. We start by the vector normal to the shock $\mathbf{\hat{n}}$. To get the normal to the shock, we use the bow shock model by \citeA{farris1991thickness}. By determining where the MMS spacecraft crosses the model bow shock boundary we can calculate the local normal to the model shock surface. There are various methods one can use to calculate $\mathbf{\hat{n}}$, either methods relying on the timing of the observation of the shock between the four spacecraft, methods relying on local measurements, or methods based on a global model of the bow shock \cite{schwartz1998}. The first method requires large spacecraft separation so the time shift between the different measurements would be observable. For most of our events, the separation between the MMS spacecraft is small, $90 \; \%$ of the events have average spacecraft separation less than 40 km. Although this is enough to resolve time shifts necessary to capture local variations of the shock surface, it is not large enough to resolve the time shift necessary to determine the global normal of the shock. As for the second method, it requires the determination of an upstream and a downstream interval on which one applies the coplanarity theorem to calculate the shock normal \cite{schwartz1998,abraham1976interplanetary}. For this method to work the upstream and downstream intervals should be far enough from the ramp so the magnetohydrodynamic (MHD) description of the shock, underlying the method, would hold. Below we describe a method to automatically separate upstream and downstream parameters to calculate the compression ratios. Those parameters can be used to calculate the normal to the shock, although this method is expected to work for quasi-perpendicular shocks, things become difficult for quasi-parallel shocks where upstream plasma parameters can be highly affected by the shock itself or be taken from the foreshock region instead of the upstream solar wind. In such cases, determination of the normal to the shock using this method becomes less reliable. Hence, we use the model bow shock method to determine the shock normal.

As mentioned in the previous paragraph, using local measurement for the plasma parameters could be problematic, especially for quasi-parallel shocks, where there is often no exact upstream/downstream transition in the local measurement due to the extended foreshock. For that reason, to calculate the main shock parameters we use time-shifted data from spacecraft located upstream of MMS, provided by the OMNI database \cite{king2005solar}. Of those shock parameters we mention the Alfv\'{e}nic Mach number in the normal incidence frame, $M_A = \frac{\mathbf{V_u}\cdot \mathbf{n}}{V_A}$ with $V_A$ is the Alfv\'{e}n velocity, the fast mode Mach number in the normal incidence frame, $M_f = \frac{\mathbf{V_u}\cdot\mathbf{n}}{V_f}$ with $V_f$ is the fast mode velocity, the angle between the upstream magnetic field and the shock normal $\theta_{Bn}$ and the upstream plasma beta $\beta$. For each of the crossings, we get the necessary quantities (magnetic field, velocity, density, and temperature) for a time interval of 10 minutes centered around the time of the shock crossing. We make sure that the interval contains measurement for more than 50\% of the interval, and then average the quantities to obtain one upstream measurement to calculate the above mentioned shock parameters. Furthermore, OMNI database does not provide an electron temperature measurement which is necessary for the calculation of $M_f$, a typical value of 12.06 eV was used instead \cite{1996newbury}. For events where OMNI data is not available, we use a value of $-10^{30}$ as a fill value for the parameter. For 349 events in the database, one or more of the parameters obtained from OMNI data are not available.

Furthermore, the magnetic field ($\mathbf{B}$) in the solar wind could experience large variation, either in magnitude or direction, which will cause uncertainty on all parameters that require $\mathbf{B}$ to be calculated. In the database, we provide the mean value of the magnitude of the upstream magnetic field, along with its standard deviation, and the maximum angle that $\mathbf{B}$ makes with its mean direction in the 10-minutes interval. On top of that, we evaluate $M_A$, $M_f$  and $\theta_{Bn}$ throughout the 10 minutes interval using $\mathbf{B}$, the standard deviation of both quantities in that interval is taken as an error estimate. In addition to these parameters, we include the upstream velocity, density, ion temperature, magnetic field vector, magnetic field magnitude, and the solar wind dynamic pressure for each shock crossing. 

\begin{figure*}[ht]
    \centering
    \includegraphics[scale=0.3]{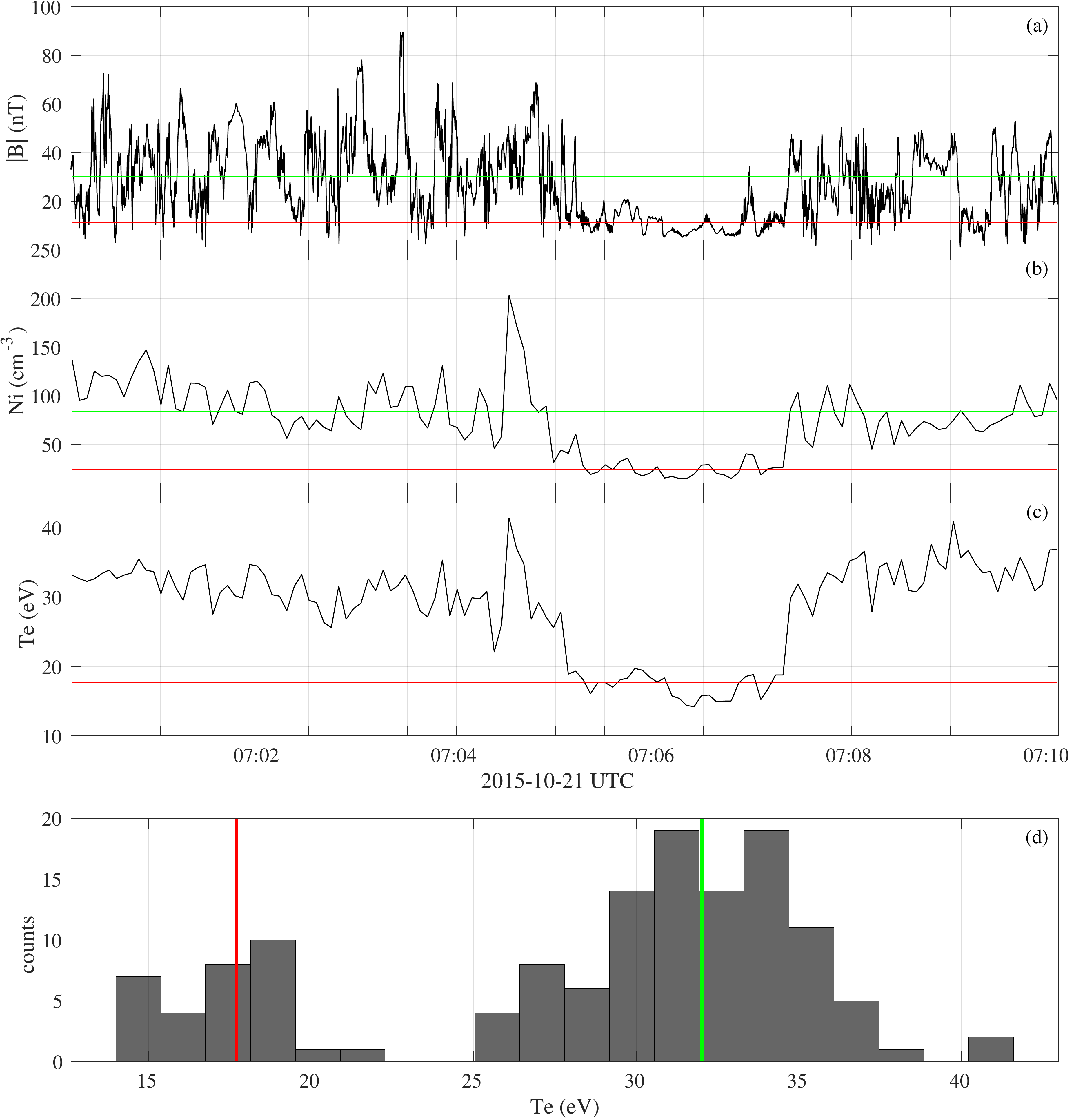}
    \caption{(a) Magnetic field magnitude, (b) ion density, (c) electron temperature and (d) histogram of the electron temperature data. Green and red horizontal lines in panels (a--c) and vertical lines in panel (d) represent the determined downstream and upstream values respectively.}
    \label{fig:fig2}
\end{figure*}

Knowledge of the downstream shock parameters is essential for the determination of various quantities, such as the compression ratios. To get those values we can use the probability output from the CNN to find intervals of magnetosheath around a shock crossing. As mentioned before, in mixed regions, i.e foot and foreshock, the CNN is not able to specify with high confidence what region MMS is in, so using the CNN probabilities to get an estimate of the downstream shock parameters can be inaccurate. To calculate those parameters with better accuracy, for each shock crossing we plot the histogram of the fast mode electron temperature data measured by FPI. Because of the shock transition, the values will be mostly separated into two distributions corresponding to the upstream and the downstream intervals. This is shown in Figure \ref{fig:fig2}, where panels (a--c) show the magnetic field magnitude, ion density and electron temperature respectively, while panel (d) shows the histogram of the electron temperature data, for a quasi-parallel shock crossing at $\theta_{Bn} \sim 33^\circ$ and $M_A \sim 9.4$. It is clear from panel (d) that we have two separate distributions, one corresponding to the upstream values (left) and one for the downstream (right). 
We can also determine the distributions in the magnetic field and the ion density corresponding to the same distribution from the electron temperature. Once we have the two different distributions we calculate its median to get a value for the upstream and downstream parameters. The green and red horizontal lines in panels (a--c) indicate the obtained median downstream and upstream values for each parameter for this event. Using those upstream and downstream values we report the B field, ion density, and electron temperature compression ratios.

Moving to the parameters related to the spacecraft and data acquisition. There are three data acquisition modes on board of MMS: slow survey, fast survey, and burst, where slow has the lowest resolution and burst has the highest \cite{fuselier2016magnetospheric}. Due to the limited telemetry rate on board of MMS, not all captured burst data can be sent to Earth. Only limited scientifically relevant periods will be selected to have data at the burst acquisition rate.
Those regions are marked either automatically through an onboard data quality value, or manually through Scientist-In-The-Loop (SITL), where scientists look at the survey mode data to determine regions of interests\cite{fuselier2016magnetospheric}. It is of interest for scientists to know if burst data exist for a certain event since much more science can be explored with such intervals. Hence, for each crossing in the database, we check if there is burst data within an interval of $\pm 5$ minutes around the crossing time. The entries in the database named ``burst\_start" and ``burst\_end" provides the start and end times of the burst interval available for each shock crossing. If no burst interval exists, the values are set to zero. 

For each shock crossing, we also include the location of the spacecraft in the Geocentric Solar Ecliptic (GSE) coordinate system in km, the spacecraft separation, and the spacecraft formation, all of which is information that could be useful while studying collisionless shocks. The spacecraft separation is quantified by the entry ``sc\_sep" in the database containing: $(\Delta R)_\mathrm{min}$ the minimum separation, $(\Delta R)_\mathrm{max}$ the maximum separation and $\langle \Delta R \rangle$ the average separation between the four spacecraft. As for the spacecraft formation, we use the tetrahedral quality factor, defined in \citeA{fuselier2016magnetospheric}, which measures how close is the formation of the spacecraft is to a tetrahedron.
A summary of the different entries in the database along with a short description is provided in Table \ref{tab:signif_gap_clos}.

\begin{figure*}[ht]
    \centering
    \includegraphics[scale=0.55]{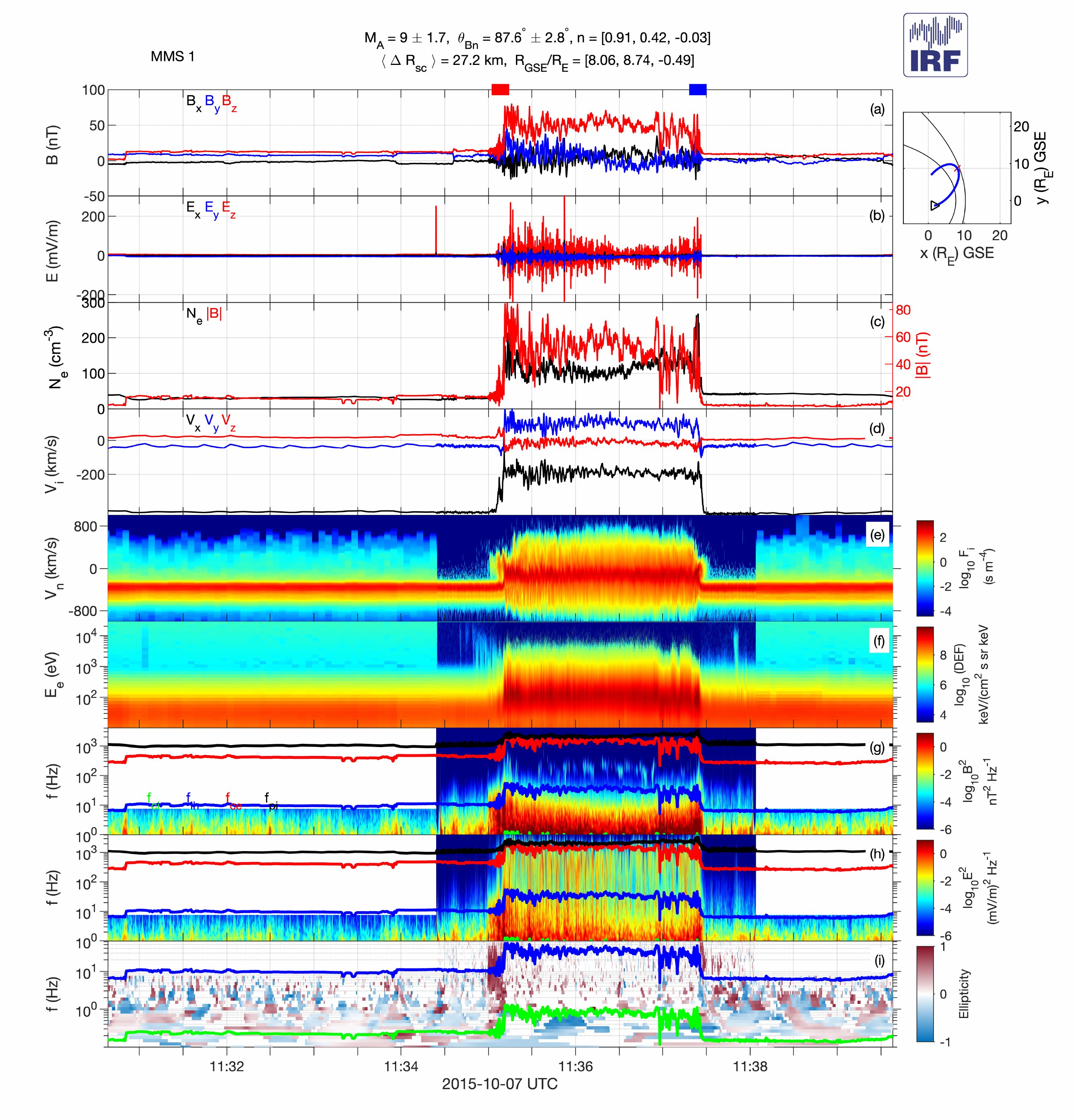}
    \caption{Overview plot example showing two quasi-perpendicular shock crossings. (a) Magnetic field, (b) electric field, (c) electron density (black) and magnetic field magnitude (red), (d) ion velocity, (e) ion velocity distribution function reduced in the normal direction, (f) electron differential energy flux, (g) magnetic field power spectral density, (h) electric field power spectral density and (i) ellipticity}
    \label{fig:fig3}
\end{figure*}

Finally, for each shock crossing, we provide an overview plot containing essential information about the shock and the nearby plasma environment. An example overview plot is shown in Figure \ref{fig:fig3}. Panels (a--b) show the magnetic field and the electric field, panel (c) shows both the electron density in black and the magnetic field magnitude in red, panel (d) shows the ion velocity, panels (e--f) show the ion velocity distribution function reduced in the direction of the normal to the shock and the omnidirectional electron differential energy flux respectively, panels (g--h) show the magnetic field and the electric field power spectral density respectively, and finally panel (i) shows the ellipticity of the magnetic field for frequencies where the degree of polarization is larger than 0.7 calculated using singular value decomposition (SVD) \cite{santolik2003}. Panels (f--i) have the ion and electron cyclotron frequencies (green and red), the lower hybrid frequency (blue), and the ion plasma frequency (black) overlaid. All vector quantities are in the GSE coordinate system. Furthermore, for each figure, we include the spacecraft with which the measurement was made and the key information about the shock crossing: $M_{A}$, $\theta_{Bn}$ along with their uncertainties, the shock normal in GSE, average spacecraft separation, and the vector location of the spacecraft in GSE and units of $R_E$. At the top of each figure we mark in red the location of the current shock crossing, and in blue other shock crossings in the plotted interval that are included in the database as well. We also include a plot showing the location of the spacecraft at the time of crossing in the ecliptic plane along with the trajectory of MMS in an interval of $\pm 10$ hours. The triangle marks the start of the orbit. On top of that we overlay a model bow shock and magnetopause. In each figure, we plot a 10 minutes interval centered around the shock crossing time in the database, and we use both fast and burst mode data overlaid on top of each other whenever the latter is available.

\begin{figure*}[ht]
    \centering
    \includegraphics[scale=0.55]{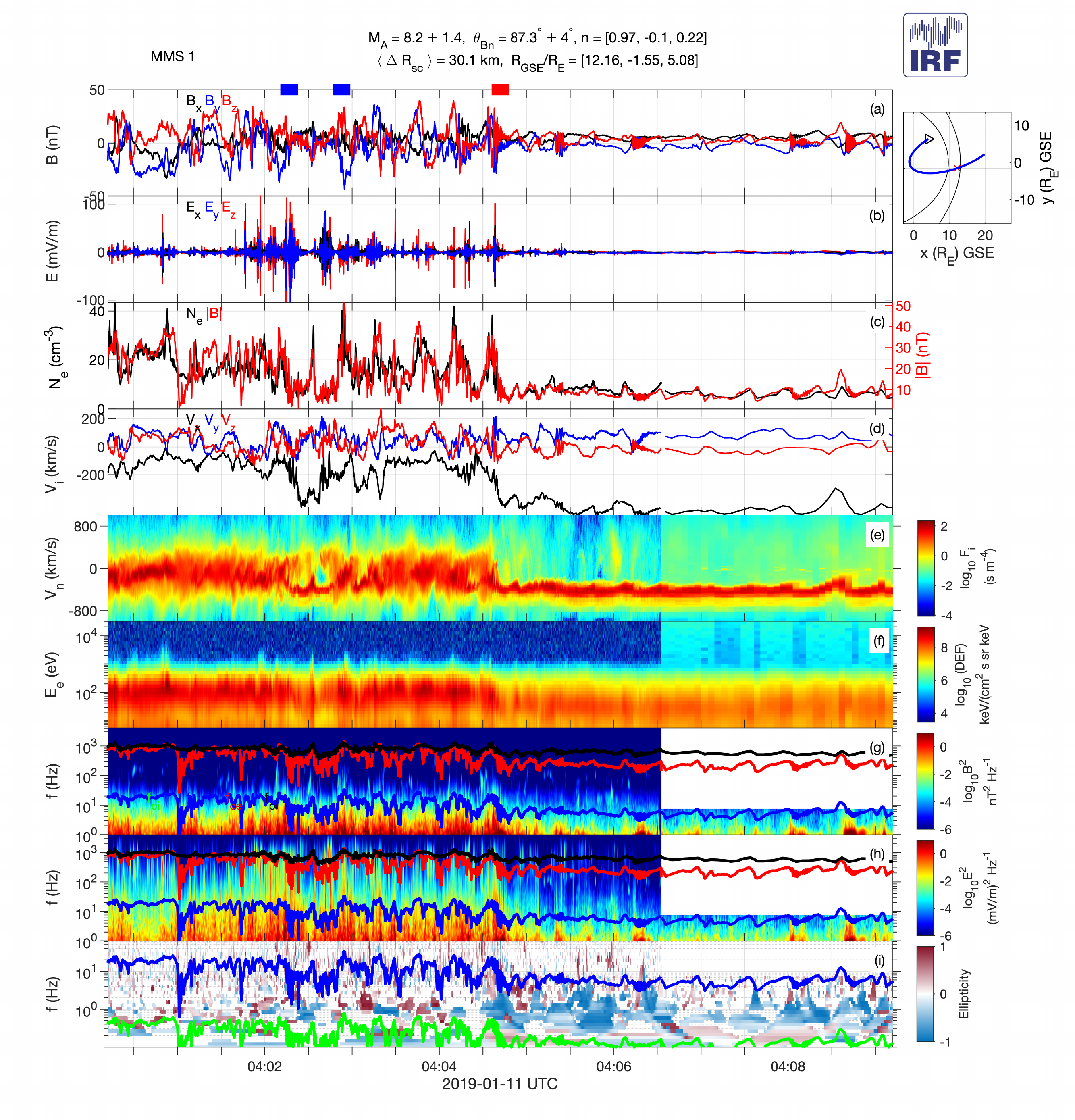}
    \caption{Overview plot example showing a quasi-parallel shock crossing misidentified as a quasi-perpendicular crossing. Same format as Figure~\ref{fig:fig3}.}
    \label{fig:fig4}
\end{figure*}

\begin{figure*}[ht]
    \centering
    \includegraphics[scale=0.55]{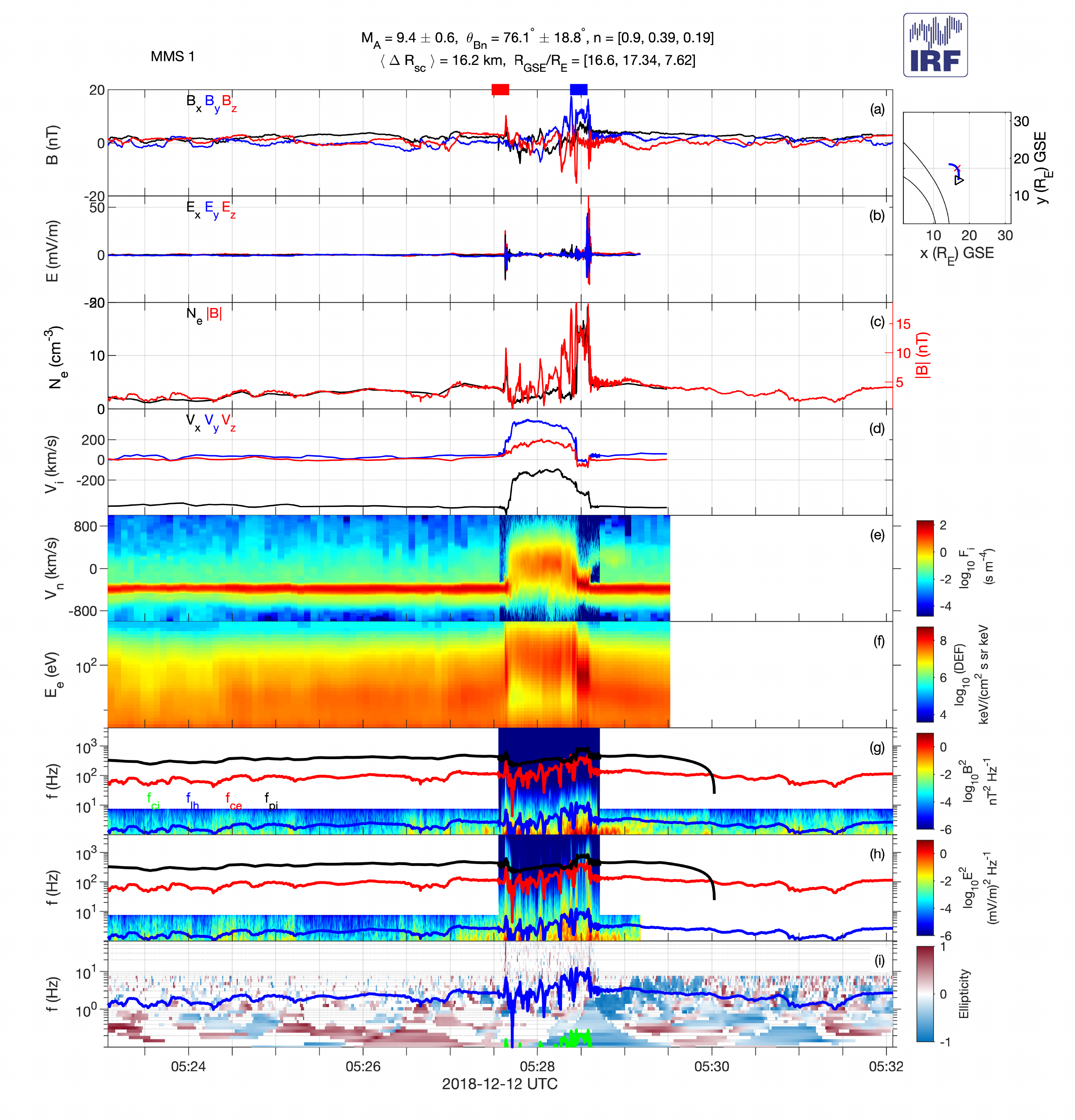}
    \caption{Overview plot example showing a hot flow anomaly identified as two shock crossings. Same format as Figure~\ref{fig:fig3}.}
    \label{fig:fig5}
\end{figure*}

We end this section by mentioning some caveats. First, multiple crossings of the same shock are included as separate shock crossings as is shown in Figure \ref{fig:fig3}. Furthermore, since we use OMNI data for calculating the shock parameters, a mismatch between the values of the parameters calculated and the expected values from observation can occur. An example of that is shown in Figure \ref{fig:fig4}, which, using the OMNI data and spacecraft position resulted in $\theta_{Bn} = 87.3^\circ$. However, the high-energy ions around the shock and the turbulent upstream and downstream signify a quasi-parallel shock. If we calculate the shock normal of this event using the mixed-mode 3 method [equation (10.17) in \citeA{schwartz1998}] and using local upstream and downstream measurements we get $\theta_{Bn} = 23^\circ$. Finally, it is worth noting that some foreshock structures, like hot flow anomalies, were identified as shock crossings by the CNN since they constitute a crossing from unshocked to shocked plasma, we kept them in the database since it's not straightforward to differentiate them from partial shock crossings without analyzing the events in detail. An example of such a case is shown in Figure \ref{fig:fig5}. In the following section, we will show that such caveats are not numerous and the information provided in the database is generally reliable.

\section{Statistics and possible application}
\label{section:4}

\subsection{Statistics}
\begin{figure*}[ht]
    \centering
    \includegraphics[scale=0.3]{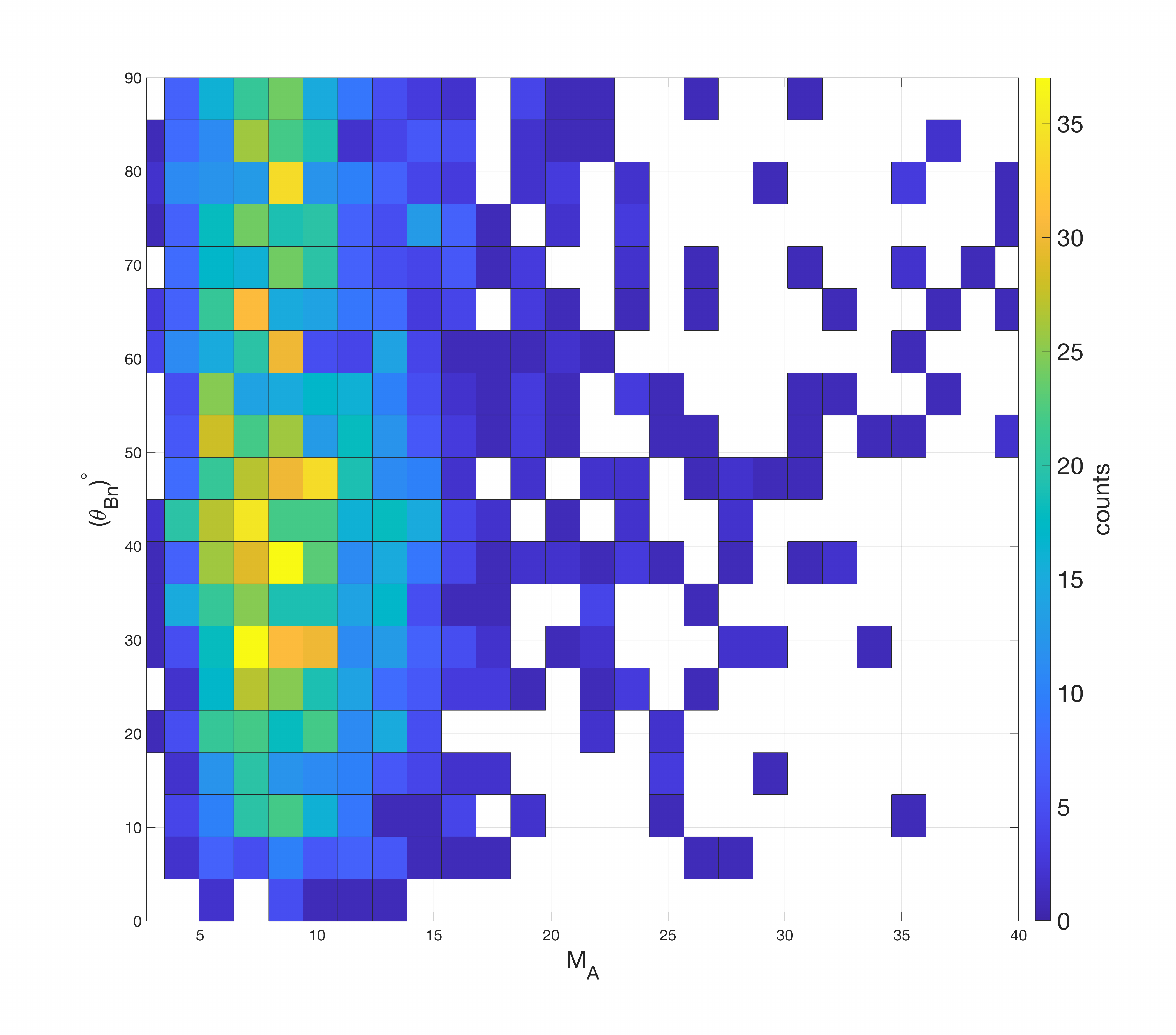}
    \caption{Distribution of the shock crossings in the database in $\theta_{Bn}$--$M_{A}$ space.}
    \label{fig:fig6}.
\end{figure*}

In compiling this database we tried to minimize human intervention as much as possible not to bias the database in parameter space. To see how the shocks are distributed in parameter space we first plot a 2D histogram in $\theta_{Bn}$ - $M_{A}$ space shown in Figure \ref{fig:fig6}, where the colorbar represent the event count in each bin. We see that the shocks cover the range in $\theta_{Bn}$  almost evenly with 45.5\% of the shocks being quasi-parallel ($\theta_{Bn}<45^\circ$), 51.4\% of the shocks being quasi-perpendicular ($\theta_{Bn}>45^\circ$) and it was not possible to compute $\theta_{Bn}$ for the remaining 3.1\% . Furthermore, we see that the shocks cover a large range of Mach numbers with the highest counts between $M_A =$ 5 and 15 which are the typical Mach number values for the solar wind as calculated from OMNI for the period between 1995 to 2018 \cite{johlander2019ion}. In this plot, we limit the Mach number range to 40 but there are entries where the Mach number exceeds this. There are some cases where the Mach number is around 150, and such shocks are associated with a very low upstream magnetic field. This causes the Mach number to become very high, but this also makes $M_{A}$ sensitive to small variations in B, and therefore these shocks typically have large uncertainties on their parameters.

\begin{figure*}[ht]
    \centering
    \includegraphics[scale=0.3]{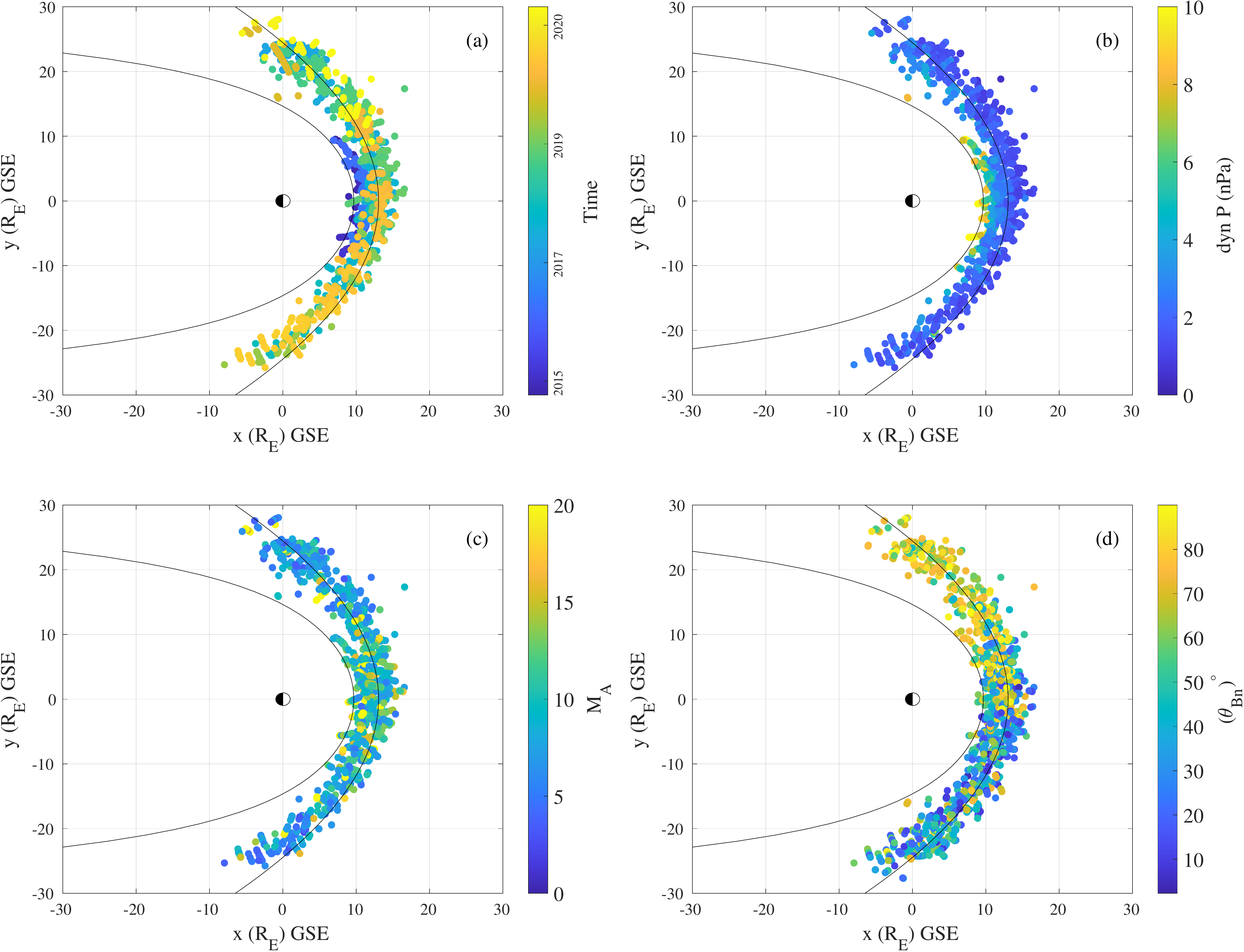}
    \caption{Location of the crossings of all of the shocks in the database projected on the ecliptic plane and normalized to earth radius. Colorbar in panel (a) represents time, (b) dynamic pressure, (c) $M_A$ and (d) $\theta_{Bn}$ with the y coordinate of the crossing flipped if the magnetic field is in the ortho-Parker spiral configuration.}
    \label{fig:fig7}
\end{figure*}

It is of interest to see how the physical locations of the shock crossings are distributed around Earth and how they are related to different parameters. Figure \ref{fig:fig7} shows the location of all the crossings in the database projected on the ecliptic plane and normalized to the Earth radius. We overlay a model magnetopause \cite{shue1998magnetopause} and bow shock \cite{farris1991thickness} whose locations are calculated using the dynamic pressure $P = 2.9$ nPa and $B_z = -0.29$ nT, averaged over all shocks in the database. The colorbar in each panel represents a different quantity, panel (a) shows the time of the crossing, (b) dynamic pressure from OMNI, (c) $M_{A}$, and (d) shows $\theta_{Bn}$ .

The first point to note from Figure~\ref{fig:fig7} is that the shocks cover a large spatial range from the subsolar point reaching the flanks at $y \sim \pm 30 R_E$. Furthermore, looking at panel (a) we see that in the early phase of the mission, before 2017, the shock crossings were closer to Earth due to the lower apogee of MMS's orbits  of 12 $R_E$ during that phase of the mission (Phase 1), which focused on the dayside magnetopause. In that same year, MMS entered Phase 2 of the mission where the apogee was raised in discrete steps to 25 $R_E$\cite{fuselier2016magnetospheric}. Crossings detected in Phase 1 of the mission are expected to occur at high dynamic pressure conditions since the magnetosphere has to be compressed to a large extent to reach MMS orbit. This is seen in panel (b) where the solar wind dynamic pressure is the highest for the shocks closest to the Earth. Panel (c) shows no particular pattern for the distribution of Mach number with the locations of the crossings. All panels are in the GSE coordinate system except for panel (d) where to account for the ortho-Parker spiral configuration of the interplanetary magnetic field (IMF) (a configuration where the IMF is at an angle of $90^\circ$ to the Parker spiral) we invert the sign of the y coordinate of the shock crossings. The purpose of this adjustment is to maintain the general trend, related to the Parker spiral, that the dusk flank is quasi-perpendicular and the dawn flank is quasi-parallel, which is visible in panel (d). This also highlights events where $\theta_{Bn}$ calculated from the OMNI database does not reflect the geometry of the shock, as discussed before and as seen in Figure \ref{fig:fig4}. Figure \ref{fig:fig7} (d) shows that those events are not frequent in the database. It is worth noting that the point in the upper right side of the plots, with maximum $x \sim 17 R_E$ is the same event shown in figure \ref{fig:fig5}.
\begin{figure*}[ht]
    \centering
    \includegraphics[scale=0.3]{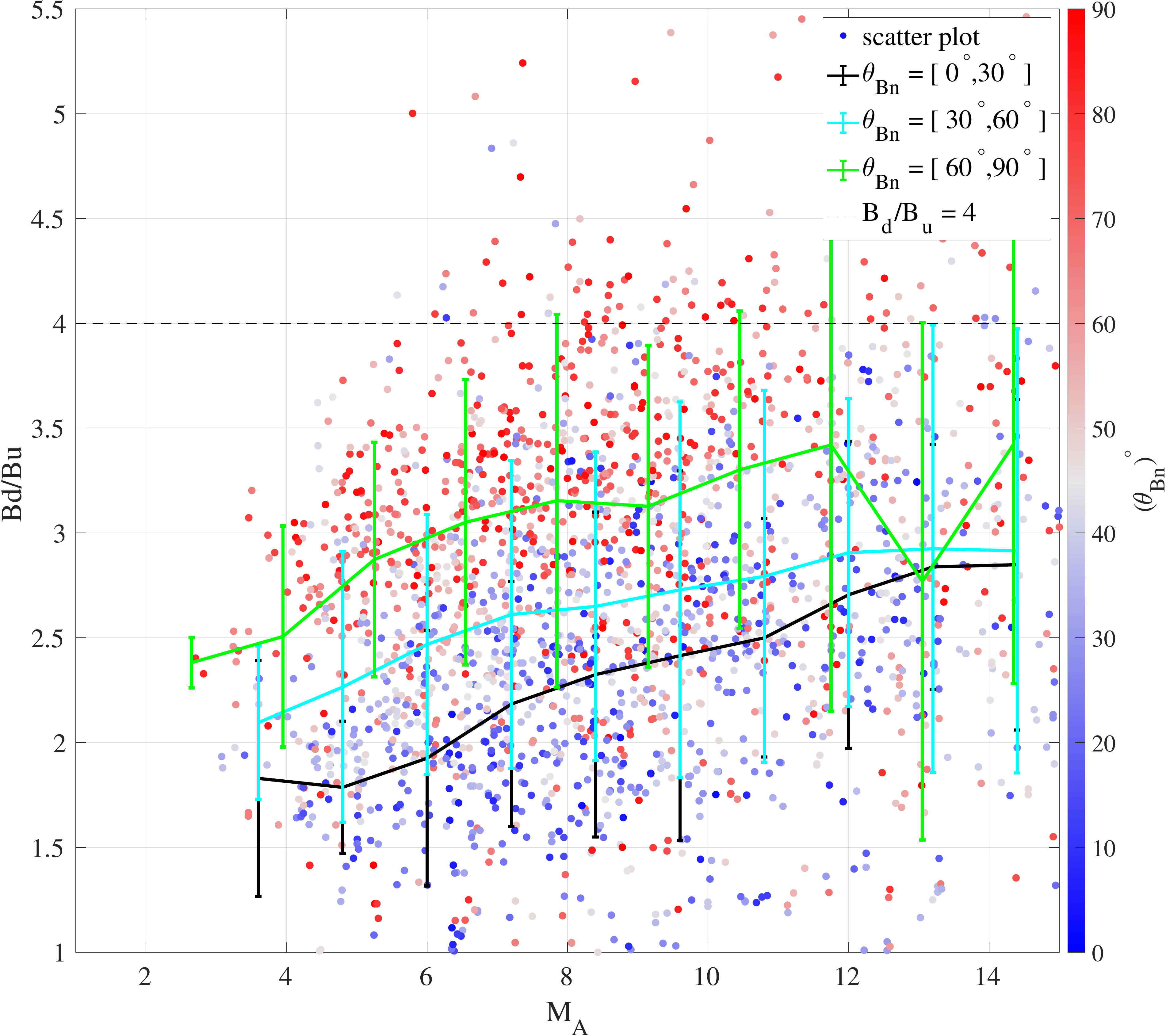}
    \caption{Magnetic field compression ratio versus the Alfv\'{e}n Mach number. Errorbars show the spread of $B_d/B_u$ for various intervals of $\theta_{Bn}$. The dashed line represent the canonical high-Mach-number limit of 4. The color shows $\theta_{Bn}$. }
    \label{fig:fig8}.
\end{figure*}

Finally, we explore how well the calculated compression ratios match expectations from Rankine-Hugoniot jump conditions. By solving the Rankine-Hugoniot conditions, we can obtain a relation between the compression ratios and the different shock parameters. In particular, in Figure \ref{fig:fig8} we plot the magnetic field compression ratio versus $M_A$ for various $\theta_{Bn}$. Comparing our result to the simulation result shown in Figure 4 in \citeA{kennel1985quarter} we see that we retrieve a similar trend where the more perpendicular the shock is, the higher the compression ratio becomes. For large $M_A$ the compression ratio approaches the expected asymptotic value of 4.  

\subsection{Statistical study of ion acceleration efficiency}
Our database can be used for a variety of applications such as identifying events with given parameters on which case studies can be conducted, or performing statistical studies. In particular, it can be interesting to study shocks that correspond to a particular parameter range, such as quasi-parallel or quasi-perpendicular shocks geometries. Furthermore, one could be interested in comparing in situ observations with remote sensing observations, i.e comparing shocks in the heliosphere with astrophysical shocks, for that comparison to be valid the shocks have to be close in parameter space. In both examples having a database that allows filtering of events with various parameters such as $M_{A}$ and $\theta_{Bn}$ is of great use. Moreover, the whole database forms the backbone for whatever statistical study that one wishes to do, either by using all of the events in the database or by selecting a subset of it. The database also increases the efficiency for many shock studies since the initial time consuming task of identifying suitable events is reduced.

To demonstrate the usability of the database, we now employ the database to study energetic ions at the bow shock. This was recently tackled by \citeA{johlander2021ion} from MMS with a set of 154 shock crossings, but here we can investigate this with a number of shocks that is over an order of magnitude larger. We calculate the ion acceleration efficiency defined as
\begin{equation}
    \epsilon (E_0) = \left< \frac{U_i(E_i>E0)}{U_i(E_i>0)} \right>,
\end{equation}
where $U_i(E_i>E0)$ is the ion energy density downstream of the shock in the local plasma frame above the threshold energy $E_0$ expressed as
\begin{equation}
    U_i(E_i>E_0) = 4\pi \sqrt{\frac{2}{m_i^2}} \int_{E_0}^{E_{max}} dE_i \sqrt{E_i^3} f_i(E_i).
\end{equation}
We set $E_0$ to 10 times the solar wind energy \cite{caprioli2014simulations,johlander2021ion}.
We use the ion distribution functions measured by FPI \cite{pollock2016fast} on MMS to calculate the acceleration efficiency, and for this, we use only the downstream distributions. We determine the downstream velocity using the same method which we have used for the determination of the compression ratios (see Section \ref{section:3}). Then we use the obtained velocity to transform the observed ion distributions to the plasma frame in the downstream region. Some events have such high solar wind speed, that the energy density calculation is done based only on two energy bins in the distribution function. For such events, the acceleration efficiency calculations are not reliable, so we remove them from the dataset. In addition, accounting for the events where there is no OMNI data, we are left with 2384 shock crossings, 53\% of which are quasi-perpendicular while the remaining 47\% are quasi-parallel. 

\begin{figure*}[ht]
    \centering
    \includegraphics[scale=0.32]{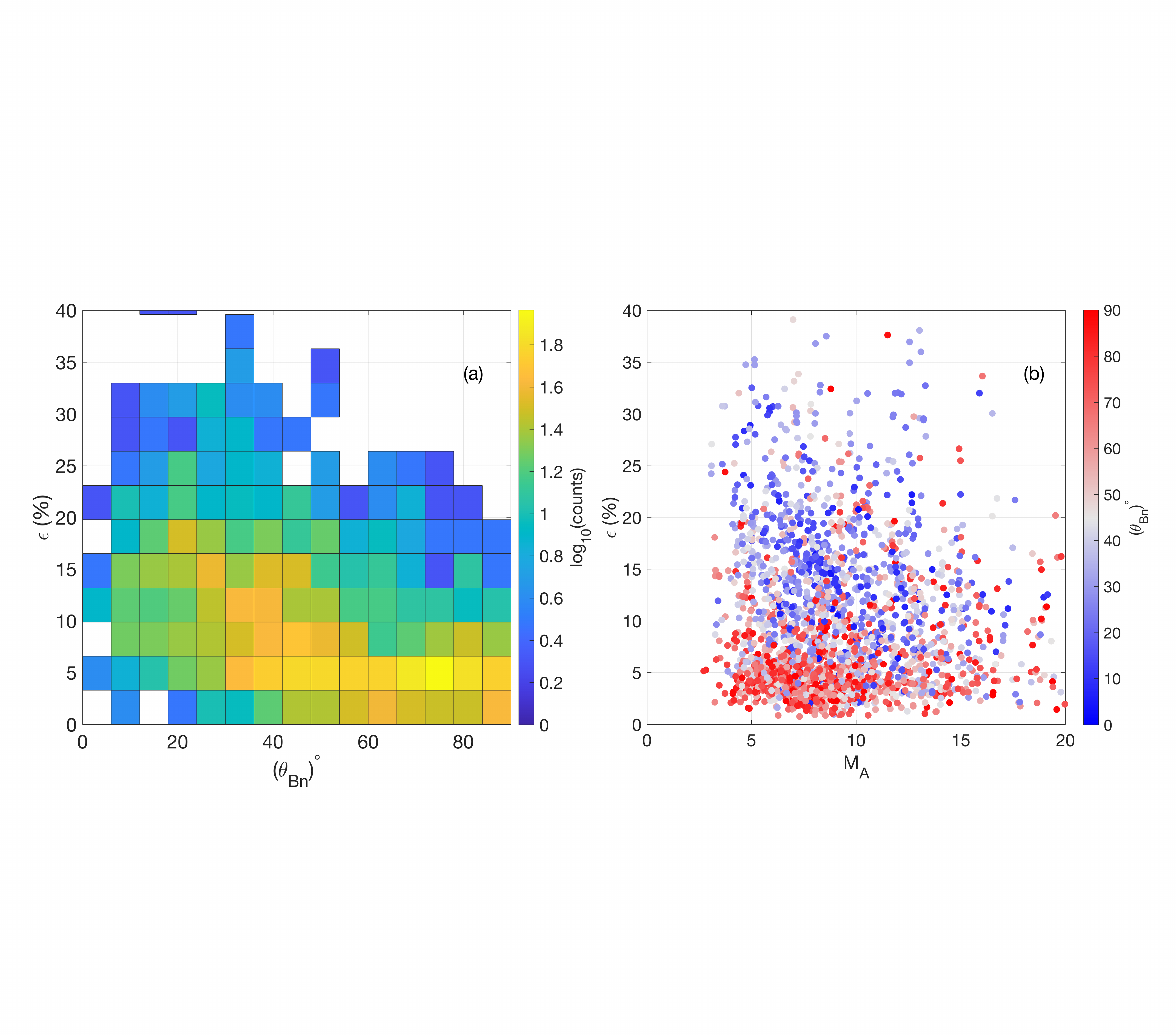}
    \caption{Ion acceleration efficiency. (a) 2D histogram of $\epsilon$ versus $\theta_{Bn}$ with color bar representing log$_{10}$ of counts. (b) scatter plot of $\epsilon$  versus $M_A$ with the colorbar showing $\theta_{Bn}$. }
    \label{fig:fig9}
\end{figure*}
The resulting ion acceleration efficiency $\epsilon$ is shown in Figure \ref{fig:fig9}, where panel (a) shows a 2D histogram in the $\epsilon$ - $\theta_{Bn}$ space with the colorbar representing the base ten logarithm of counts for each bin and panel (b) shows a scatter plot of $\epsilon$ versus $M_A$ where the colorbar represents $\theta_{Bn}$ where the color scale has been set so that blue represents events with $\theta_{Bn}<45^\circ$ while red represent events with $\theta_{Bn}>45^\circ$. From panel (a) one can clearly see the higher spread and average value of $\epsilon$ for quasi-parallel shocks compared to quasi-perpendicular ones. If we calculate the weighted mean and standard deviation for all quasi-parallel and all quasi-perpendicular shocks, taking the count in each bin of the histogram as a weight we get $\langle \epsilon \rangle = 14 \pm 11 \%$ for the former and $8 \pm 8 \%$ for the latter. This shows that quasi-parallel shocks are more efficient at accelerating ions compared to quasi-perpendicular shocks.

These results are in agreement with those of \citeA{johlander2021ion} who also found that quasi-parallel shocks are much more efficient at accelerating ions than quasi-perpendicular ones, where the acceleration efficiency increased at $\theta_{Bn}<50^\circ$ . Also, they observe that $\epsilon$ decreases for $\theta_{Bn}<20^\circ$, where this decrease is attributed to a low number of events in that range. Having a larger sample size we don't observe the same decrease at low $\theta_{Bn}$. Furthermore, \citeA{johlander2021ion} observed a dependence of the acceleration efficiency on Mach number where lower $M_A$ events have lower $\epsilon$, a trend not present in the current data set (see Figure~\ref{fig:fig9} (b)). 

This statistical study was performed on the database as is, as mentioned earlier, events with $\theta_{Bn}$ values that do not reflect the geometry of the shock and events with foreshock structures labeled as shock crossings were included in this statistics. As shown earlier, such events are not numerous, and should not affect the results, nevertheless in a future dedicated study this will be carefully considered and such events will be filtered out.

\section{Conclusion} 
\label{section:5}

In this report, we present a database of Earth's bow shock crossings by MMS spacecraft compiled using a machine learning algorithm. We show that shock crossing events can be reliably identified using automated methods with little bias in parameter space. The database contains 2797 shock crossing events, the largest bow shock crossing database so far, covering a broad range in parameter space, as well as in physical space with crossings from the sub-solar point towards the flanks. For each crossing time, we provide key information related to each shock such as the Alfv\'{e}nic Mach number or $\theta_{Bn}$. We also provide an overview plot for each event showing the most important quantities related to the observed shock. This database will be a large asset to the community, facilitating statistical studies and case studies of single events.

We demonstrate a use for the database by performing a quantitative study of ion acceleration efficiency at the bow shock. Using a dataset of 2000+ shocks, we show that quasi-parallel shocks are more efficient at accelerating ions than quasi-perpendicular shocks in agreement with the result of \citeA{johlander2021ion}. We also show that there is no correlation between the ion acceleration efficiency and $M_A$ in contrast to the results of \citeA{johlander2021ion}, which shows the advantage of having a database that is comprised of more events.

\begin{table}
    \caption{Parameters included in the shock database. All vector quantities are in the GSE coordinate system. All plasma and field measurement, except for the compression ratios, are from the OMNI database.}
    \label{tab:signif_gap_clos}
        \begin{tabular}{p{0.2\linewidth} p{0.6\linewidth} p{0.2\linewidth}}
        \hline
        Parameter name in DB & Description & Units  \\
        \hline

        time  & Date and time interval of the shock crossing & Unix Epoch (in seconds since 1 January 1970)\\
        direction & Flag indicating if the shock is inbound ($1$) or outbound ($-1$)  & - \\
        burst\_start - burst\_end & Start and end times of the burst interval if available zero if not available & Unix Epoch (in seconds since 1 January 1970)\\
        Bx\_us - By\_us - Bz\_us & upstream magnetic field vector $[B_x,B_y,B_z]$  & nT\\
        B\_us\_abs & upstream magnetic field magnitude & nT\\
        sB\_us\_abs & Standard deviation on the magnitude of the upstream magnetic field & nT\\
        Delta\_theta\_B & The maximum rotation of the upstream magnetic field vector in the OMNI interval used & degrees\\
        Ni\_us & Upstream ion density  & cm$^{-3}$\\
        Ti\_us & Upstream ion temperature & eV\\
        Vx\_us - Vy\_us - Vz\_us & Upstream velocity  & km/s\\
        beta\_i\_us & Upstream ion $\beta$ & - \\
        Pdyn\_us & Upstream dynamic pressure from OMNI & nPa\\
        thBn & Angle between upstream magnetic field and shock normal & degrees\\
        sthBn & Standard deviation of  $\theta_{Bn}$ based on variation in upstream B in the OMNI interval used  & degrees\\
        normal\_x - normal\_y - normal\_z & Shock normal $\mathbf{n}$ from \citeA{farris1991thickness} model of the bow shock & - \\
        MA & Alfv\'{e}nic Mach number in the normal incidence frame assuming a stationary shock & -\\
        sMA & Standard deviation of MA based on variation in upstream B in the OMNI interval used & - \\
        Mf & Fast mode Mach number in the normal incidence frame assuming a stationary shock with $T_e = 12.06 eV$ & -\\
        sMf & Standard deviation of Mf based on variation in upstream B in the OMNI interval used & - \\
        B\_jump & Magnetic field compression ratio & -\\
        Ni\_jump & Ion density compression ratio & -\\
        Te\_jump & Electron temperature ratio & -\\
        pos\_x - pos\_y - pos\_z & The location of the spacecraft when it observed the shock & km\\
        sc\_sep\_min - sc\_sep\_max - sc\_sep\_mean & Spacecraft separation $[\min \; \max \;\mathrm{mean}]$ separation & km\\
        TQF & Tetrahedral quality factor measuring how close the MMS formation is to a tetrahedron & - \\
        \hline
    \end{tabular}
\end{table}

\acknowledgments
The database and the overview plots can be found at https://doi.org/10.5281/zenodo.6343989. We thank the entire MMS team and instrument PIs for data access and support. This database is part of the EU Horizon 2020 SHARP project, which will include shocks from multiple spacecraft and from different locations in the heliosphere. MMS data are available at \url{https://lasp.colorado.edu/mms/sdc/public/data/} following the directories: mms\#/fgm/brst/l2 for FGM data, mms\#/fpi/brst/l2/dis-dist for FPI ion distributions, mms\#/fpi/brst/l2/dis-moms for FPI ion moments and mms\#/fpi/brst/l2/des-moms for FPI electron moments. OMNI data used are available at \url{https://omniweb.gsfc.nasa.gov/}. Data analysis was performed using the IRFU-Matlab analysis package available at \url{https://github.com/irfu/irfu-matlab}. This work is supported by the Swedish Research Council grant 2018-05514 and the European Union's Horizon 2020 research and innovation program under grant agreement number 101004131 (SHARP). APD received financial support from the Swedish National Space Agency (Grant  number 2020-00111)

\bibliography{biblio}

\end{document}